\RequirePackage{ifpdf}
\ifpdf 
\documentclass[pdftex]{sigma}
\else
\documentclass{sigma}
\fi

\begin{document}

\renewcommand{\PaperNumber}{002}
\FirstPageHeading

\ShortArticleName{Multi-Instantons and Superstring Solitons}

\ArticleName{Multi-Instantons in Higher Dimensions\\ and
Superstring Solitons}

\Author{Eugene K. LOGINOV}

\AuthorNameForHeading{E.K. Loginov}

\Address{Department of Physics, Ivanovo State University,
 39 Ermaka Str., Ivanovo, 153025 Russia}
\Email{\href{loginov@ivanovo.ac.ru}{loginov@ivanovo.ac.ru}}

\ArticleDates{Received June 30, 2005; Published online August 16, 2005}

\Abstract{We construct octonionic multi-instantons for the eight
and seven dimensional Yang--Mills theory. Extended soliton
solutions to the low-energy heterotic field theory equations of
motion are constructed from these octonionic multi-instantons. The
solitons describe a string in ten-dimensional Minkowski space, and
preserve one and two of the sixteen space-time supersymmetries
correspondingly.}


\Classification{83E30}

\section{Introduction}

We search solutions for lowest nontrivial order in $\alpha'$
of the equations of motion that follow from the bosonic action
\begin{gather*}
S=\frac{1}{2k^2}\int d^{10}x\,\sqrt{-g}e^{-2\phi}
\left(R+4(\nabla\phi)^2-\frac{1}{3}H^2-\frac{\alpha'}{30}\,\text{Tr}\,F^2\right).
\end{gather*}
We are interested in solutions that preserve at least one
supersymmetry. This requires existence in ten dimensions of
at least one Majora\-na--Weyl spinor $\epsilon$ such that the
supersymmetry variations of the fermionic fields vanish for such
solutions
\begin{gather}
\delta\chi=F_{MN}\Gamma^{MN}\epsilon,\nonumber\\
\delta\lambda=\left(\Gamma^{M}\partial_{M}\phi-\frac16H_{MNP}\Gamma^{MNP}\right)\epsilon,\nonumber\\
\delta\psi_{M}=\left(\partial_{M}+\frac14\Omega_{M}^{AB}\Gamma_{AB}\right)\epsilon,
\label{2}
\end{gather}
where $\chi$, $\lambda$ and $\psi_{M}$ are the gaugino,
dilatino and the gravitino fields, respectively.
The gene\-ra\-lized spin connection $\Omega$ is a non-Riemannian
connection related to the spin connection $\omega$
and the anti-symmetric tensor field strength $H$ by
\begin{gather*}
\Omega_{M}^{AB}=\omega_{M}^{AB}-H_{M}^{AB}.
\end{gather*}
In \cite{1}
 a one-brane solution of heterotic theory
 which is an everywhere smooth solution of the equations of motion was found.
 Construction of this solution involves essentially properties
 of octonions. One of the many bizarre features of
 this soliton is that it preserves only one of
 the sixteen space-time supersymmetries,
 in contrast to previously known examples
 of supersymmetric solitons which all preserve half
 of the supersymmetries.
 In \cite{2} a two-brane solution of heterotic theory was found.
 This soliton preserves two of the sixteen supersymmetries
 and hence corresponds to $N=1$ space-time supersymmetry
 in $(2+1)$ dimensions transverse to the seven dimensions where
 the Yang--Mills instanton is defined.

\section{Multi-instantons in eight dimensions}

We start by picking a particular commuting $SO(8)$ spinor $\eta$ with
$\Gamma_9\eta=\eta$ normalized to $\eta^{T}\eta=1$. Then we can
introduce a fourth-rank antisymmetric $Spin(7)$-invariant tensor
\begin{gather*}
f_{mnps}=-\eta^{T}\Gamma_{mnps}\eta.
\end{gather*}
There exists an explicit construction of the $SO(8)$ gamma
matrices in terms of the octonion structure constants $c_{ijk}$
defined by
\begin{gather*}
e_{i}e_{j}=-\delta_{ij}+c_{ijk}e_{k},
\end{gather*}
where $c_{ijk}$ are antisymmetric in $(i,j,k)$, non\-zero and
equal to unity for the seven combinations $(123)$, $(145)$,
$(167)$, $(246)$, $(275)$, $(374)$, $(365)$. Using this
construction and an explicit choice for~$\eta$ can find
\begin{gather*}
f_{ijk8}=c_{ijk},\qquad
f_{ijkl}=\delta_{il}\delta_{jk}-\delta_{ik}\delta_{jl}+c_{ijr}c_{klr}.
\end{gather*}
Suppose $e_{ps}$ are the standard generators of the Lie algebra
$so(8)$. Define elements of the subalgebra $so(7)$ of $so(8)$ by
\begin{gather*}
E_{mn}=\frac18(3\delta_{mp}\delta_{ns}-3\delta_{ms}\delta_{np}-f_{mnps})e_{ps}.
\end{gather*}

Consider now the Yang--Mills gauge theory in eight dimensions with
the gauge group $Spin(7)$. We proceed from the following ansatz
\begin{gather*}
A_m=\frac{4}{3}\frac{\lambda^{\dag}y^{n}}{(1+y^{\dag}y)}E_{mn},
\end{gather*}
where $y$ is a column vector with the octonions $y_1,\dots,y_{N}$
such that
\begin{gather*}
y^{\dag}=(y^{k}_1,\dots,y^{k}_{N})\bar e_{k},\qquad y^{k}_{I}\in{\mathbb R},\nonumber\\
\lambda^{\dag}=(\lambda_1,\dots,\lambda_{N}),\qquad \lambda_{I}\in{\mathbb  R}^+,\nonumber\\
y^{k}_{I}=(b^{k}_{I}+x^{k})\lambda_{I}.
\end{gather*}
Using the switching relations for the generators $E_{mn}$, we get
the self-dual field strength
\begin{gather*}
F_{mn}=-\frac{4}{9}\lambda^{\dag}\left[\frac{3(2+2y^{\dag}y-y^{i}y_{i}^{\dag})
E_{mn}+(3\delta_{mi}\delta_{ns}-3\delta_{ms}\delta_{ni}-f_{mnis})
E_{sj}y^{j}y_{i}^{\dag}}{(1+y^{\dag}y)^2}\right]\lambda.
\end{gather*}
Thus we have a $Spin(7)$-invariant solution of the Yang--Mills
field equations which depends on at most $9N$ effective
parameters~(cf.~\cite{1}).

\section{Superstring solitons}

Now we denote world indices of the eight-dimensional space
transverse to the 1-brane by $\mu,\nu=1,\dots, 8$
and the corresponding tangent space indices by $m,n=1,\dots, 8$.
We assume that no fields depend on the longitudinal coordinates
and that the nontrivial tensor fields in the solution have
only transverse indices. Then the gamma matrix
terms in \eqref{2} are sensitive only to the
$Spin(7)$ part of $\epsilon$.
Thus taking $\epsilon$ to be a $Spin(7)$ singlet $\eta$ and the
non-vanishing components of $F_{MN}$ to be those given by the
eight dimensional octonionic multi-instanton the supersymmetry
variation $\delta\chi$ of the gaugino vanishes. This follows from
the fact that
\begin{gather*}
(3\delta_{mp}\delta_{ns}-3\delta_{ms}\delta_{np}-f_{mnps})\Gamma^{ps}\eta=0
\end{gather*}
and self-duality of $F_{\mu\nu}$. To deal with the other
supersymmetry variations, we must adopt an ansatz for the
non-trivial behavior of the metric and antisymmetric tensor fields
in the eight dimensions transverse to the 1-brane. Let
\begin{gather*}
g_{\mu\nu}=e^{(6/7)\phi}\delta_{\mu\nu},\qquad
H_{\mu\nu\lambda}=\frac{1}{7}f_{\mu\nu\lambda\sigma}\partial^{\sigma}\phi,
\end{gather*}
where $\phi$ is to be identified with the dilaton field. With this
ansatz, we can prove that it suffices to take $\epsilon$ to be a
constant $Spin(7)$ invariant spinor to make the dilatino variation
$\delta\lambda $ and the gravitino variation $\delta\psi_{M}$
vanish.

In order to determine the solution completely, we need only to solve the
Bianchi identity
\begin{gather*}
dH=\alpha'\left(\text{tr}\,R\wedge R-\frac{1}{30}\text{Tr}\,F\wedge
F\right),
\end{gather*}
where $\text{Tr}$ refers to the trace in the adjoint
representation of $E_8$ or $SO(32)$ in the corresponding heterotic
string theory. For order $\alpha'$ we can neglect the first term
and obtain the following dilaton solution:
\begin{gather*}
e^{-(6/7)\phi}=e^{-(6/7)\phi_0}+\frac{32}{9}
\frac{\lambda^{\dag}(2+2y^{\dag}y-y^{i}y_{i}^{\dag})\lambda}{(1+y^{\dag}y)^2}.
\end{gather*}
The metric and antisymmetric fields are constructed of this dilaton
field according to the space\-time-supersymmetric ansatz. However,
the metric and the field strength fall of only as $1/r^2$ and this
implies that the ADM mass per unit length of this string diverges to be
just like the mono-instanton solution (see~\cite{1}).

\section{Multi-instantons in seven dimensions}

Consider now the Yang--Mills gauge theory in seven dimensions with
the gauge group $G_2$. This group may be defined as the group of
all automorphisms of the algebra of octonions $\mathbb O$. The
corresponding Lie algebra $g_2$ is generated by all derivations of
the form
\begin{gather*}
D_{ij}:\ z\to \frac16[[e_{i},e_{j}],z]+\frac12(e_{i},e_{j},z),
\end{gather*}
where the associator $(x,y,z)=(xy)z-x(yz)$ is skew-symmetric for
$x,y,z\in\mathbb O$. We can then introduce a completely
antisymmetric tensor $c_{ijkl}$ by
\begin{gather*}
(e_i,e_j,e_k)=2c_{ijkl}e_l.
\end{gather*}
It is easy to prove that this tensor is $G_2$-invariant. We choose
the ansatz
\begin{gather*}
A_m=\frac{3}{2}\frac{\lambda^{\dag}y^{n}}{(1+y^{\dag}y)}D_{mn},
\end{gather*}
where $y$ is the above-mentioned column vector of octonions.

Using the switching relations for the derivations $D_{mn}$, we get
the self-dual field strength
\begin{gather*}
F_{mn}=-\frac{3}{4}\lambda^{\dag}\left[\frac{2(2+2y^{\dag}y-y^{i}y_{i}^{\dag})
D_{mn}+(2\delta_{mi}\delta_{ns}-2\delta_{ms}\delta_{ni}-c_{mnis})
D_{sj}y^{j}y_{i}^{\dag}}{(1+y^{\dag}y)^2}\right]\lambda.
\end{gather*}
Thus we have a $G_2$-invariant solution of the Yang--Mills field
equations which depends on at most~$8N$ effective parameters. In
fact we obtain $E_6$-invariant solution of the equations.
Indeed, $E_6$~is a group of linear transformations of the
exceptional Jordan algebra $J(\mathbb O)$ which preserve a cubic
form (norm) $n(X)$ of any element $X\in J(\mathbb O)$. It can be
shown that there exists a set of elements
$X_{ijkl}=X(e_{i},e_{j},e_{k},e_{l})$ of $J(\mathbb O)$ such that
the norm
\begin{gather*}
n(X_{ijkl})=c_{ijkl}.
\end{gather*}
Since the group $G_{2}$ can be isomorphically enclosed into the
group $E_{6}$, we prove the $E_{6}$-invariance of the found
solution.

\section{Superstring solitons}

Let us now show that this solution can be extended to a soliton
solution of the heterotic string. To obtain this we choose $\epsilon$
to be a constant $G_2$ invariant spinor, and the metric and
antisymmetric tensor fields to be of the form
\begin{gather*}
g_{\mu\nu}=e^{\phi}\delta_{\mu\nu},\qquad
H_{\mu\nu\lambda}=\frac{1}{4}f_{\mu\nu\lambda\sigma}\partial^{\sigma}\phi.
\end{gather*}
To order $\alpha'$ we obtain the following dilaton solution:
\begin{gather*}
e^{-\phi}=e^{-\phi_0}+
\frac{4\lambda^{\dag}(2+2y^{\dag}y-y^{i}y_{i}^{\dag})\lambda}{(1+y^{\dag}y)^2}.
\end{gather*}
Obviously, this solution also does not have finite energy per unit
length, and therefore it complicates physical implications.
However, as in the mono-instanton case (see~\cite{2}), we
can suppose that this divergent energy is an infrared phenomenon
and does not preclude the existence of a~well-behaved low-energy
effective action governing the string dynamics on scales large
relative to its core size.

\begin{remark}
In eight and seven dimensions, in contrast to four dimensions, the
Yang--Mills action is infinite and the topological meaning of the
solutions is unknown. In the case we cannot assert that these
solutions are gauge equivalent to the mono-instanton ones
that were found in previous papers. They are new solutions which
depend on at most $9N$ and $8N$ effective parameters
correspondingly.
\end{remark}

\begin{remark}
The fact that these solutions do not have finite action involves
that the correspon\-ding supersymmetric solutions do not have finite
energy per unit length. This is a characteristic feature of such
solutions. There are no finite-action solutions of the
Yang--Mills equations in eight and seven dimensions. In order to
get finite-action Yang--Mills equations one would
``compactify'' some of the dimensions.
\end{remark}

\begin{remark}
These notes are based on my lecture at the Sixth International
Conference ``Symmetry in Nonlinear Mathematical Physics'' (June
20--26, 2005, Kyiv, Ukraine). More detailed version of these notes
can be found in \cite{3,4}.
\end{remark}

\subsection*{Acknowledgements}
Research supported by RFBR Grant 04-02-16324.

\LastPageEnding

\end{document}